# Performance / Price Sort

Jim Gray, Joshua Coates, Microsoft Research, Chris Nyberg, Ordinal Technologies
Gray@Microsoft.com, t-joshc@microsoft.com, Chris@Ordinal.com
April 1998, revised July 1998

**Abstract:** NTsort is an external sort on WindowsNT 5.0. It has minimal functionality but excellent price performance. In particular, running on mail-order hardware it can sort 1.5 GB for a penny. NT5.0 is not yet available. For commercially available sorts, Postman Sort from Robert Ramey Software Development has elapsed time performance comparable to NTsort, while using less processor time. It can sort 1.27 GB for a penny (12.7 million records.) These sorts set new price-performance records. This paper documents this and proposes that the PennySort benchmark be revised to *Performance/Price* sort: a simple GB/$ sort metric based on a two-pass external sort.

**Why does anyone care about sorting and sort performance?**   The prosaic reason is that sorting is a common task -- it is frequently used in database systems, data analysis, and data mining. Another important reason is that *sorting is a simple balanced workload*, involving memory access, IO, and cpu. It evaluates a computer system's overall performance. Being simple, sorting is easily ported from one system to another, easily scaled to large SMP systems, and to computer clusters.

The first public sort benchmark was defined in *A Measure of Transaction Processing Performance*, *Datamation*, April 1, 1985 [1]. That article defined **DatamationSort** to measure how fast can you sort a million records. The records are 100 bytes, with 10-byte keys in random order. The sort is external (disk-to-disk.) The time includes starting the program, creating the target file, and doing the sort. Prices are list prices depreciated over 3 years. (see http://research.microsoft.com/barc/SortBenchmark/ for the rules).

Since then, there has been steady improvement in sort performance: sort speeds have approximately doubled each year, and price performance has approximately doubled each year -- improving a thousand-fold every decade. In part, this has been due to Moore's law, things get faster every year: but that is only 40% of the story. The other 60% came from better algorithms and from parallelism. The current champion, NOWsort, used 95 UltraSPARCs to sort 8.4 GB in a minute.

DatamationSort times were getting tiny (a few seconds) and so it seemed better to have a fixed-time benchmark rather than a fixed-size sort. *MinuteSort*, how much can you sort in a minute, replaced DatamationSort in 1994. MinuteSort is a "biggest bang" (price is no object) test. *PennySort* is a "bang-for-the-buck" measure, how many 100-byte records you can sort for a penny, if the system cost is depreciated over three years.   Inexpensive systems are allowed to run for a long time while expensive systems must run the sort quickly.

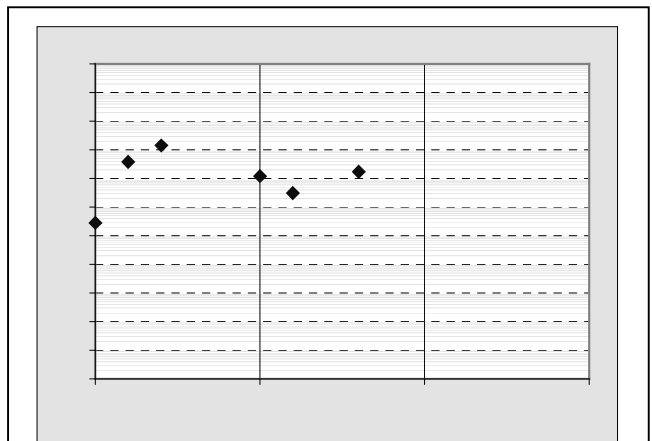

**Figure 1:** The top graph shows sort speed over the last 15 years. The bottom graph shows GB sorted per dollar. The large diamond and triangle show the NTsort results. Given the focus on price/performance, NTsort is 5x above the price trend line. NTsort uses 95x fewer processors and disks -- that's why it is 82x slower than NOWsort.

Several MinuteSort benchmark results have been reported, but until now, there have been no PennySort results. In fact, the PennySort Benchmark was originally the DollarSort benchmark. It was changed to PennySort when someone pointed out that a DollarSort would run for 50,000 seconds on a 2,000$ system. PennySort brought this time down to 500 seconds. However, that was the wrong answer.   If innovation and Moore's Law continue, in a less than a decade PennySort will be back up to the 50,000 seconds.

Consequently, this paper both reports the first PennySort result and proposes to redefine the benchmark into a **Performance/Price** sort: the $/GB cost of a two-pass external sort.

## *1998 Daytona & Indy PennySort*

Three commercial sorts are running on Windows NT 4.0.3 at our lab:
(1) NT Sort included in NT 5.0 but not yet commercially available,
(2)  NitroSort v 1.0A  from Opnek Research, Hackettstown, NJ
    (299$ from http://www.nitrosort.com/nitrsort.html) and,
(3) Postman's Sort v 3.21. . from Robert Ramey Software Development, Santa Barbara, CA
    ( 149$ from http://www.silcom.com:81/~ramey/)
Other commercial sorts include Syncsort (http://www.syncsort.com/infosnt.htm) and CoSort (http://www.iri.com/98new/win32.html).   These sorts are as expensive as the PennySort machine, and so are more appropriate for MinuteSort benchmarks.

All three sorts are single-threaded, so they run best on uni-processors. We originally estimated that a 2k$ system should be able to sort 700MB for a penny.  The system would have the fastest CPU and memory that 1.4k$ can buy, a 2 GB data source and target disk, and a 1 GB scratch disk.  It would have 64 MB of DRAM (to produce 25 25MB data runs on the scratch disk using 256KB writes) and then merge them into the target (using 256KB reads and writes). 500$ was budgeted for NT Workstation and the sort software.

We were aware of the substantial price differential between SCSI and IDE disk drives.  Traditionally, the SCSI price was justified because IDE did Programmed Input-Output (PIO) rather than Direct Memory Access (DMA).  PIO moves each byte through the CPU registers rather than having the host-bus adapter stream the data over the bus directly to memory via DMA.  Recently, IDE drives adopted UltraDMA IDE - - the IDE adapter does not interrupt the processor during a disk transfer.   Consequently, the cpu overhead for an IO goes from 6 instructions per byte transferred under PIO to 0.2 instructions per byte transferred under DMA.  This is a dramatic saving, without DMA the system has 80% processor utilization when transferring 8 MBps.  With DMA, the utilization drops to 2%.  Therefore, UltraDMA IDE drives compete with SCSI for small disk arrays. SCSI is still advantageous for strings of disks: IDE does not allow multiple outstanding commands on one string.

Shopping the web found an mail-order house that would sell us the system described in Table 1. These are OEM prices.  These items are not individually available at this price (for example, the best individual price of NT Workstation is 140$).   The prices are here to show the cost breakdown in the pie chart at right.

| Table 1: Price breakdown of PennySort Machine. www.pricewatch.com | | |
|---|---|---|
| quantity | part | price |
| 1 | Assembly | 25 |
| 1 | ASUS P2L97Motherboard + Intel P2-266 MHz + fan + 1 Year Warranty | 495 |
| 1 | 64MB SDRAM (10ns) | 94 |
| 1 | Mini Tower Case w/235W Power Supply & fan | 47 |
| 1 | floppy drive | 18 |
| 2 | Fujitsu MPS3032UA 3.1 GB U/ATA 10ms 128k 5400rpm | 278 |
| 1 | INTEL 8465 EtherExpress PCI Pro/100 | 48 |
| 1 | Virage 3D Graphics card w/2MB EDO | 33 |
| 1 | NT Workstation 4.0.3 Software | 69 |
| 1 | Shipping (3 day) | 35 |
| | **Total:** | **1142** |

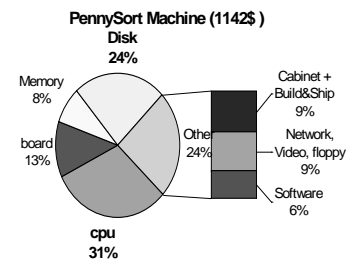

The key properties of this system are Intel Pentium II 266 MHz processor, Asus mother board, 64 MB of Synchronous 10 ns SDRAM, dual 3.5GB Fujitsu UltraDMA IDE drives, and NT Workstation, all for $1,142. Given this price, the time allowed for a PennySort, computed by Table 2, is

| **Table 2:** Computing the time budget for PennySort: SystemCost/seconds_per_3_Years | |
|---|---|
| seconds/3 years | 94,608,000 |
| PennySort HW+SW cost | $1,142 |
| seconds/penny on PennySort | **828** |

828 seconds. The prices keep dropping. Today (the end of April 1998) http://www.pricewatch.com/ reports that component prices have dropped 10% this month.

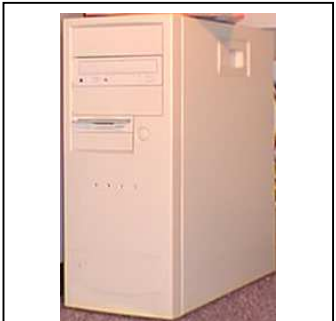

**Figure 2.** The PenySort Machine: what 1.4K$ buys in May 1998: a 266 Mhz Intel Pentium II, 64MB 10ns SDRAM, 2 UltraDMA IDE drives, Windows NT Workstation, 100 Mbps Ethernet, assembly and 1 year guarantee, 2$^{nd}$ day shipping from mail-order house. Display, mouse, and keyboard are sold separately ☺.

We first benchmarked the UtraIDE disks without installing the DMA software (the PiiXide driver, called that because it runs on the Pentium II). The non-DMA system ran 80% CPU saturated when reading or writing the disks at 8 MBps. With the PiiXide driver installed, the CPU utilization dropped 40x to 2%. The disk transfer rate is limited by the rate at which bytes move past the disk read/write heads. The disks are banded: outer bands transfer at almost 9 MBps, inner bands transfer at 6 MBps. The average rate across the surfaces is 7.8MBps.

Both disks had a small (400MB) FAT partition to store NT programs (they came from the supplier that way). We added a (3GB) NTFS partition to store the input and output data, and the sort temporary file. The data flow is as shown in Figure 2. The data moves to or from disk 4 times in a 2-pass sort. During both phases, both the input and output can be overlapped. Therefore, in theory, the sort can fully utilize the bandwidth of both disks. Doing a simple copy (using fast-unbuffered copy from the MSDN VC++ examples) showed that the IO time for the data flows of Figure 3 moving 1.45 GB is 421 seconds elapsed, 9.6 seconds of cpu time, and .4 seconds of user time. So, the sort will be IO bound if it uses less than 411 seconds of CPU time.

An August 1, 1997, *PC Week* article by John Shumate, reviewed several sort programs: *4 Programs Make NT 'Sort' of Fast* [18]. He observed that NTsort produced no output and no error if the file was larger than memory. He also pointed out that NTsort's other shortcomings: no GUI, no API, and extremely limited function. Consequently, he recommended the other sort programs, notably CoSort, Optec, Nitro, and Postman.

We repaired NTsort to be a simple two-pass sort. If the input fits in available memory, NTsort does one pass. If not, NTsort allocates memory appropriate for a two-pass sort: essentially the square-root of the input file size times the transfer size -- 20 MB in the case of a 1 GB sort. Sort then reads a block of data, quick-sorts it, and writes out the sorted run to a temporary file. After the input is converted to sorted runs, NTsort uses a tournament tree to merge the runs. NT sort does not overlap or pipeline the phase-one IO. If the input or output is a file, NTsort uses unbuffered IO.

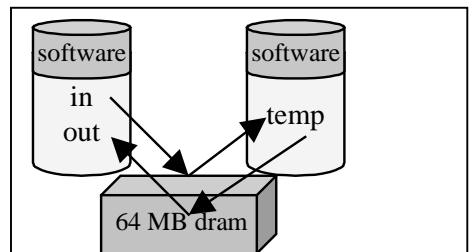

**Figure 3:** Data flow of a two-pass sort: (1) input is sorted into runs stored on the temporary file. Then (2) runs are merged to form the sorted output

The complete documentation for NTsort is given in the text box below. *Minimal* is one word that comes to mind. Anyone wanting a full function sort should look elsewhere. Both PostmanSort and NitroSort are inexpensive, well documented, and have a full-function API and command interface. NitroSort has a very nice GUI as well. Nevertheless, NTsort can run the PennySort benchmark and measure NT's IO performance.

```
SORT [/R] [/+n] [/M kilobytes] [/L locale] [/RE recordbytes]
     [[drive1:][path1]filename1] [/T drive2:][path2]] [/O [drive3:][path3]filename3]
  /+n                          Specifies the character number, n, to begin each comparison.
                               /+3 indicates that each comparison should begin at the 3rd
                               Lines with fewer than n characters collate before other lines.
  /L[OCALE] locale             Overrides the system default locale with the specified one.
                               The "C" locale is the fastest collating sequence.
                               It is currently the only alternative.
                               The sort is always case insensitive.
  /M[EMORY] kilobytes          Specifies amount of main memory to use for the sort, in kilobytes.
  /REC[ORD_MAXIMUM] characters Specifies the maximum number of characters in a record
                               (default 4096, maximum 65535).
  /R[EVERSE]                   Reverses the sort order; that is, sorts Z to A, then 9 to 0.
  [drive1:][path1]filename1    Specifies the file to be sorted. If not specified, the standard input is sorted.
                               Specifying the input file is faster than redirecting the same file as standard input.
  /T[EMPORARY] [drive2:][path2]    Specifies the path of the directory to hold the sort's working storage,
                               in case the data does not fit in main memory.
                               The default is to use the system temporary directory.
  /O[UTPUT][drive3:][path3]filename3   Specifies the file where the sorted input is to be stored.
                               If not specified, the data is written to the standard output.
                               Specifying the output file is faster than redirecting standard output to the same file.
```

We generated a 15 M record (1.45 GB) file using the SortGen program [23]. We sorted the result using NT Sort (with "C" locale), Postman Sort, and NitroSort. We then normalized the results within the time-budget computed by dividing the system price by the 3-year depreciation. The results are shown in Table 3. NitroSort 1.0.a limits itself to 4 MB by default. For a 1.5GB sort, that implies a 3-pass sort and much longer run times. So, we gave NitroSort a hint: telling it to use 21 MB of memory. It then ran in reasonable time. Both PostmanSort and NitroSort use the NT File system buffered IO (rather than direct IO). That explains why they have much larger kernel times. On the other hand, they use much less user time, so they use 2x to 4x less cpu time than NTsort, showing that they have much better sort algorithms. Since sort is IO bound, this better cpu performance is not reflected in the elapsed time. Subsequently, Jay Cole of Openk Research gave us access to NitroSort 1.5. It improved cpu and elapsed times

| Table 3. 1997 PennySort times. | | | | | | | |
|---|---|---|---|---|---|---|---|
| Product | Time Budget | Kernel | User | Total cpu time | Sorted MB | GB/$ | Category |
| NTsort | 828 | 7 | 402 | 409 | 1,445 | 141 | Indy |
| PostmanSort | 733 | 61 | 151 | 212 | 1,277 | 125 | Daytona |
| NitroSort (no hint) | 656 | 22 | 27 | 49 | 355 | 35 | Daytona |
| NitroSort hint | 656 | 40 | 60 | 99 | 727 | 71 | Daytona |
| NitroSort 1.5 Beta | 656 | 53 | 60 | 113 | 980 | 96 | Post April 1 |

NTsort and PostmanSort have comparable running times. The extra cost of Postman Sort (149$) reduces its time budget and so it can sort less per penny than NTsort even though PostmanSort is just as fast and uses much less cpu time. Still, **Postman sort is a commercial product so it is the winner of the Daytona category** (commercially available sort programs). **NTsort wins in the Indy category** (since it has slightly better price performance but is not commercially available).

All these sorts are IO bound. If they aggressively overlapped IO and computation, they would run in about 411 seconds -- more than twice as fast. At that speed, PostmanSort and NitroSort would still use only 50% of the processor. To match the cpu speed, the PennySort machine needs disks that are 2x faster or 2-disk file striping. NTsort is not overlapping computation and IO, the other sorts are counting on the NT file system to do the overlap. This costs them about 50 seconds of cpu time (in `memcpy()`) and gives sub-optimal in IO performance.

The point of this exercise was to show how inexpensive commodity hardware can perform IO intensive tasks. The obvious next steps to improve the performance of these sorts are (1) fix the IO design to overlap computation and IO, (2) use multi-threading to exploit SMPs when the disk bandwidth is beyond the power of a single processor (about 20MBps), and (3) port to a cluster and replicate the NOWsort work (they sorted 8.4GB on 95 nodes in 60 seconds). A cluster of 32 PennySort machines using an 100Mbps Ethernet switch would cost about 25x less than the NOW cluster and should be able to sort about 10GB in a minute (about 313 MB for each node).

## *Datamation Sort Results*

We did a DatamationSort (one million records) with the three sorting programs. We added 64 MB to the PennySort machine to allow a one-pass sort of 100 MB. Postman Sort had the fastest sorting time at about 26 seconds. It performed equally well with a memory hint and without. NTSort came in second at about 35 seconds. NTSort is too conservative with its default memory allocation; on the 128 MB PennySort machine, NTSort does not default to allocate the 105 MB it needs for a one-pass sort. However, when NTSort is given a memory hint, it performs adequately. NitroSort 1.5 Beta came in third at about 55 seconds. We were unable to coax NitroSort into performing a single pass sort, even with a memory hint. The ratio between CPU and IO wait is similar to the two-pass experiments for all three sorts.

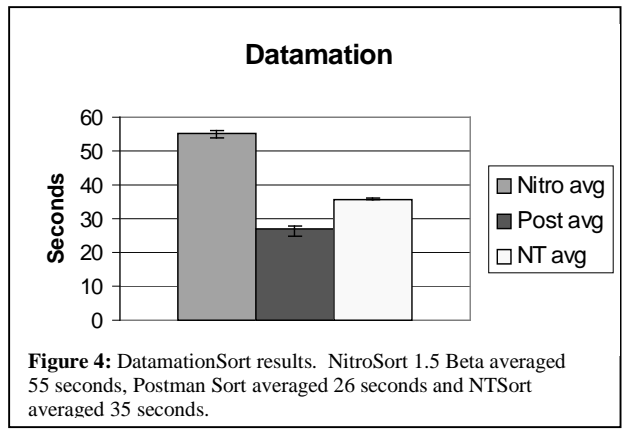

**Figure 4:** DatamationSort results. NitroSort 1.5 Beta averaged 55 seconds, Postman Sort averaged 26 seconds and NTSort averaged 35 seconds.

## *Performance/Price Sort*

PennySort is a poor benchmark definition. The idea of a fixed-price benchmark is not workable when price-performance doubles every year. You can do 1,000 times more a decade later for the same price. In particular, the PennySort of the year 2008 is likely to sort 1.5 TB and run for 800,000 seconds (over two days) on a one-dollar computer!

The goal is to have a simple and portable I/O benchmark that measures the system's performance/price. A simple way out of this is to use the two-pass minute-sort benchmark, but just aim for minimum price measured in GB/$ sorted. To be specific:
(1) Sort the largest file you can in a minute.
(2) Compute the system price per second (3-year depreciation => system price/9.5e-7 ).
(3) Compute the GB/$ sorted by dividing item 1 by item 2.
As with the current sort, there should be two categories: Daytona (commercially available general-purpose product and Indy (research hacks allowed, not necessarily a product).

The virtue of picking a minute is that it gives direct comparability with MinuteSort performance and price performance.

The current Performance/Price results are shown in Table 4:

| **Table 4:** Historical Performance/Price results. | | | | | | |
|---|---|---|---|---|---|---|
| year | MB/sec | GB/$ | System [reference] | Sys price (M$) | CPUs | |
| 1985 | 0.02 | 0.05 | M6800 Bitton et al [7,8] | 0.03 | 1 | Datamation |
| 1986 | 0.03 | 0.01 | Tandem Tsukerman [19,20] | 0.3 | 3 | Datamation |
| 1987 | 3.85 | 0.05 | Cray YMP, Weinberger [21] | 7.0 | 1 | Datamation |
| 1991 | 14.29 | 0.54 | IBM 3090, DFsort/Saber | 2.5 | 1 | Datamation |
| 1990 | 0.31 | 0.15 | Kitsuregawa [12] | 0.2 | 1 | Datamation |
| 1993 | 1.20 | 0.11 | Sequent, Graefe [11] | 1.0 | 32 | Datamation |
| 1994 | 1.72 | 0.16 | IPSC/Wisc DeWitt [10] | 1.0 | 32 | Datamation |
| 1994 | 11.11 | 5.25 | Alpha, Nyberg [7] | 0.2 | 1 | Datamation |
| 1995 | 28.57 | 2.70 | SGI/Ordinal, Nyberg [16] | 1.0 | 16 | Minute/Daytona |
| 1995 | 19.61 | 37.10 | IBM, Agarwal [2] | 0.05 | 1 | Minute/Indy |
| 1996 | 100.00 | 15.76 | NOW, Arpaci-Dusseau [3] | 0.6 | 32 | Minute/Indy |
| 1997 | 140.17 | 8.41 | Now 95 , Arpaci-Dusseau [3] | 2.0 | 95 | Minute/Indy |
| 1997 | 86.21 | 6.27 | SGI/Ordinal, Nyberg [17] | 1.3 | 14 | Minute/Datona |
| 1998 | 1.74 | 125.00 | PostmanSort | 0.0013 | 1 | Penny/Datona |
| 1998 | 1.74 | 144.00 | NTsort | 0.0012 | 1 | Penny/Indy |